\documentclass[prd,twocolumn,reprint,preprintnumbers,nofootinbib,superscriptaddress,longbibliography]{revtex4-1}

\usepackage{soul}
\usepackage{eqnarray}
\usepackage{amssymb}
\usepackage{enumerate}
\usepackage{amssymb}
\usepackage{amsmath}
\usepackage{mathtools}
\usepackage{tikz}
\usepackage{makecell}
\usepackage{natbib}
\usepackage{graphicx}
\usepackage{tensor}
\usepackage{physics}
\usepackage{slashed}
\usepackage{pifont}
\usepackage{bbold}
\usepackage{cleveref}
\crefname{equation}{Eq.}{Eqs.} 
\crefrangelabelformat{equation}{(#3#1#4--#5#2#6)}

\usepackage{slashed}
\usepackage{tikz} 
\usetikzlibrary{shapes.misc}

\newcommand{\genericT}{\ensuremath{T}}
\newcommand{\met}{\slashed{E}_\genericT}

\newcommand{\mptvec}{\slashed{\vec{p}}_\genericT}

\begin{document}

\title{When the Machine Chimes the Bell:\\
Entanglement and Bell Inequalities with Boosted $t\bar{t}$}

\author{Zhongtian Dong}\email{cdong@ku.edu}
\affiliation{Department of Physics and Astronomy, University of Kansas, Lawrence, Kansas 66045, USA}
\author{Dorival Gon\c{c}alves}\email{dorival@okstate.edu}
\affiliation{Department of Physics, Oklahoma State University, Stillwater, Oklahoma, 74078, USA}
\author{Kyoungchul Kong}\email{kckong@ku.edu}
\affiliation{Department of Physics and Astronomy, University of Kansas, Lawrence, KS 66045, USA}
\author{Alberto Navarro}\email{alberto.navarro\_serratos@okstate.edu}
\affiliation{Department of Physics, Oklahoma State University, Stillwater, OK, 74078, USA}

\begin{abstract}
The Large Hadron Collider provides a unique opportunity to study quantum entanglement and violation of Bell inequalities at the highest energy available today. In this paper,  we will investigate these quantum correlations with top quark pair production, which represents a system of two-qubits. 
The semileptonic top pair channel has a factor of $6$ increase in statistics and easier reconstruction with respect to the dileptonic channel. 
Although measuring the spin polarization of the hadronic top quark is known to be challenging, our study indicates that it is feasible to reconstruct the spin density matrix of the two-qubit system using an optimal hadronic polarimeter. This is achieved with the aid of jet substructure techniques and NN-inspired reconstruction methods, which improve the mapping between subjets and quarks. We find that  entanglement can already be observed at more than the $5\sigma$ level with existing data, and violation of Bell inequalities may be probed above the $4\sigma$ level at the HL-LHC with 3~ab$^{-1}$ of data.
\end{abstract}

\maketitle    

\section{Introduction}
\label{sec:intro}

Quantum theory provides a robust framework for describing physical phenomena at the microscopic level, and its implications are ubiquitous in our daily life. However, various aspects of quantum theory are highly nontrivial. 
One such phenomenon is entanglement, which sets quantum physics apart from classical physics~\cite{Einstein:1935rr,schrodinger_1935}. Even more perplexing is the violation of Bell  inequalities~\cite{Bell:1964kc}, a consequence of entanglement that unequivocally identifies this departure from classical theory. Since the birth of quantum theory, entanglement and violation of Bell  inequalities have been  extensively tested in a myriad of experiments, ruling out the classical hidden variable explanation of quantum phenomena. Typically, such tests are performed using entangled low-energy photons~\cite{Freedman:1972zza,Aspect:1982fx,PhysRevLett.89.240401}, ions \cite{PMID:11236986}, superconducting systems \cite{PMID:19779447}, and nitrogen vacancy centers \cite{Pfaff_2012}, among others. In recent decades, a series of experiments have closed all potential loopholes in these tests~\cite{Hensen:2015ccp,PhysRevLett.115.250401,PhysRevLett.115.250402}.

Although these  quantum correlations have been probed in the low-energy regime, they have yet to be observed in the very high-energy scales. 
Some proposals suggested testing Bell inequalities in $e^+e^-$ collisions~\cite{Tornqvist:1980af}, decays of charmonium~\cite{Baranov:2009zza,Baranov:2008zzb,Chen:2013epa}, and positronium~\cite{Acin:2000cs}.
The Large Hadron Collider (LHC) provides an excellent environment for studying quantum entanglement
and Bell inequalities at the highest energy available today. Only very recently, a few proposals have been made to test entanglement and Bell inequalities for the final state with $t \bar t$ \cite{Afik:2020onf,Fabbrichesi:2021npl,Severi:2021cnj,Aguilar-Saavedra:2022uye,Afik:2022kwm,Afik:2022dgh,Severi:2022qjy,Aoude:2022imd}, a pair of weak bosons \cite{Fabbrichesi:2023jep,Barr:2021zcp,Barr:2022wyq,Aguilar-Saavedra:2022mpg,Aguilar-Saavedra:2022wam}, and $\tau^+\tau^-$ \cite{Altakach:2022ywa,Fabbrichesi:2022ovb}.

In this paper, we explore the possibility of observing  entanglement and violation of Bell inequalities in top quark pair production, which represents a system of two-qubits. Previous studies have mainly focused on the dilepton channel~\cite{Afik:2020onf,Fabbrichesi:2021npl,Severi:2021cnj,Aguilar-Saavedra:2022uye}, as the top quark spin  is correlated with the direction of charged leptons. 
They show that entanglement may be observed for both the $t \bar t$ threshold and the boosted region, while violation of Bell inequalities may be probed only in the boosted region. 
However, the dilepton channel suffers from lower statistics and the reconstruction of two top quarks is challenging. Hence, we will consider the semileptonic channel, which will increase the signal statistics by a factor of $6$ and allow an easier reconstruction. We will address some technical challenges associated with this analysis, which can be overcome using an optimal hadronic polarimeter in a combination of jet substructure and machine learning techniques. 

Whereas the observation of entanglement and Bell inequalities is interesting in its own right, it can further shed light on top quark physics. Top quark physics is one of the most relevant topics within the standard model (SM) and in the search for new physics. Ultimately, studying entanglement and violation of Bell inequalities may offer a new perspective on these searches.
Examples include searches for new physics in top-quark pair production through new resonances~\cite{Maltoni:2024tul} and standard model
effective field theory (SMEFT) contributions \cite{Aoude:2022imd}.

This paper is organized as follows. We provide a brief review of entanglement and Bell inequalities in Sec.~\ref{sec:theory}. We then study the possibility of observing  entanglement and violation of Bell inequalities for the top quark production in the semileptonic channel. We present the parton-level truth in Sec.~\ref{sec:parton} and present the complete analysis accounting for parton-shower, hadronization, and detector effects in Sec.~\ref{sec:analysis}. Finally, Sec.~\ref{sec:summary} is reserved for summary and outlook. 

\section{Theoretical Framework}
\label{sec:theory}

\subsection{Quantum entanglement and Bell inequalities}

A quantum state of two subsystems $A$ and $B$ is separable when its density matrix $\rho$ can be expressed as a convex sum 
\begin{align}
    \rho=\sum_i p_i\rho_A^i \otimes \rho_B^i\,,
\end{align}
where $\rho^i_{A}$ and $\rho^i_{B}$ are quantum states of the subsystems $A$ and $B$, and $\Sigma_i p_i = 1$ with $p_i\ge 0$. If the state is not separable, it is named \emph{entangled}. The Peres-Horodecki criterion provides a necessary and sufficient condition for entanglement in two-qubit systems~\cite{Peres:1996dw,Horodecki:1997vt}. It exploits the fact that, for a separable density matrix $\rho$, the partial transpose for a second subsystem
\begin{align}
    \rho^{T2}=\sum_i p_i\rho_A^i\otimes (\rho_B^i)^T\,,
\end{align}
results in a non-negative operator. Therefore, if $\rho^{T2}$ displays at least one negative eigenvalue, the density matrix $\rho$ describes an entangled system.

An entangled quantum state can result in the  violation of Bell-type inequalities~\cite{Bell:1964kc}. Remarkably, the violation of such inequalities demonstrates that there is no local hidden variable theory capable of encoding the generated entanglement. Hence, quantum mechanics cannot be described by classical laws. 
The Clauser, Horne, Shimony, and Holt (CHSH) inequality is a realization of the Bell-type inequalities for bipartite systems~\cite{Clauser:1969ny}
\begin{align}
    |\langle A_1B_1\rangle +\langle A_2 B_1\rangle+\langle A_1 B_2\rangle -\langle A_2B_2\rangle|\le 2\,,
\end{align}
where the observer of system $A$ (usually called Alice) can measure two distinct observables $A_1$ or $A_2$ and the observer of system $B$ (usually called Bob) can probe $B_1$ or $B_2$. Each of these observables has two possible eigenvalues, $\pm 1$. For spin observables, the CHSH inequality can be violated by certain entangled states, provided that there is an appropriate choice of spin axes $\hat{a}_{1(2)}$ for $A$ and $\hat{b}_{1(2)}$ for $B$.

\subsection{Top polarization}
\label{sec:polarization}

In this section, we will discuss the top quark polarization and how it can be used to probe entanglement and a possible violation of the CHSH inequalities in top quark pair production $t\bar t$ at the LHC.

Because of its short lifetime $(\sim 10^{-25}~$s), the top quark decays before hadronization occurs $(\sim 10^{-24}$~s) and spin decorrelation effects take place $(\sim 10^{-21}$~s)~\cite{Mahlon:2010gw,ParticleDataGroup:2020ssz}. It implies that the top quark final states correlate with the top quark polarization axis as
\begin{align}
    \frac{1}{\Gamma}\frac{d\Gamma}{d\cos\xi_k}=\frac{1}{2}\left( 1+ \beta_k p\cos\xi_k \right)\,,
\end{align}
where $\Gamma$ is the partial decay width, $\xi_k$ is the angle between the final state particle $k$ and the top quark spin axis in the top quark rest frame, $p$ is the degree of polarization of the ensemble, and $\beta_k$ is the spin analyzing power for the decay product $k$~\cite{Bernreuther:2010ny}. The spin analyzing power $\beta_k$ is $+1$ for the charged lepton $\ell^+$ and $\bar d$-quark, $-0.3$ for the neutrino $\bar \nu$ and $u$-quark, $-0.4$ for the $b$-quark, and $0.4$ for the $W^+$ boson. The signs of $\beta_i$ coefficients are reversed for anti-top quark decays.

Similar to the charged lepton $\ell^\pm$, the $d$-quark (or down-type quark) displays maximal spin analyzing power. However, it is a challenging task to tag a $d$-quark in a collider environment. A possible solution is to choose the softest of the two light jets from the top decay in the top quark rest frame. This choice leads to a spin analyzing power  $\beta_\text{soft}\simeq 0.5$~\cite{Jezabek:1994qs,Barman:2021yfh}. This spin analyzing power can be uplifted by choosing an optimal hadronic polarimeter $\beta_\text{opt}\simeq0.64$~\cite{Tweedie:2014yda}. 

\subsection{Optimal hadronic polarimeter}
\label{sec:pola}

We begin by considering the top quark decay in the $W$ boson rest frame. 
The polar decay angle  ($\theta_{W_\text{hel}}$) in the $W$ rest frame (often called helicity angle) is strongly correlated with the polarizations of the $W$. We take the ``$z$-axis'' of the decay to be the direction pointing opposite to the $b$-quark direction, $\hat{z}=-\hat{b}$ in the $W$ boson rest frame. 

The cosine of the helicity angle $c_{W_\text{hel}} \equiv \cos \theta_{W_\text{hel}}$ is taken to be positive when the $d$-quark (or down-type quark) is emitted in the forward hemisphere and the $u$-quark (or up-type quark) in the backward hemisphere. Then, the distribution of $c_{W_\text{hel}}$ for the decay products of $W^\pm$ is given by
\begin{align}
    \rho(c_{W_\text{hel}})=\frac{1}{\sigma}\frac{d\sigma}{dc_{W_\text{hel}}}=&\frac{3}{8}f_{R}(1\pm c_{W_\text{hel}})^{2} + \frac{3}{4}f_{0}(1-c_{W_\text{hel}}^{2}) \nonumber \\
    &+ \frac{3}{8}f_{L}(1\mp c_{W_\text{hel}})^{2}.
\end{align}
The upper sign corresponds to $W^+$ and the lower sign to $W^-$. The polarization fractions can be obtained from the following relations~\cite{Bern:2011ie}
\begin{align}
f_0&=2-5\langle c_{W_\text{hel}}^{2}\rangle,\\
f_L&=-\frac{1}{2}\mp \langle c_{W_\text{hel}}\rangle+\frac{5}{2}\langle c_{W_\text{hel}}^{2}\rangle,\label{eq:fL}
\\
f_R&=-\frac{1}{2}\pm \langle c_{W_\text{hel}}\rangle+\frac{5}{2}\langle c_{W_\text{hel}}^{2}\rangle,
\label{eq:fR}
\end{align}
with the expectation of an observable $g(c_{W_\text{hel}})$ defined as
\begin{equation}
\langle g(c_{W_\text{hel}}) \rangle\equiv \int_{-1}^1 g(c_{W_\text{hel}})\frac{1}{\sigma}\frac{d\sigma}{dc_{W_\text{hel}}}dc_{W_\text{hel}}.
\end{equation}
$f_{0}={m_t^2}/{(m_t^2 + 2 m_W^2)} \approx 0.7$, $f_{L} = {(2 m_W^2) }/{(m_t^2 + 2 m_W^2)}\approx 0.3$, and $f_{R}=0$ are, respectively, the fractions of zero helicity, left-handed helicity,  and right-handed helicity of the $W^+$ boson in the top quark rest frame. For the $W^-$ decay, the quoted values of $f_R$ and $f_L$ are interchanged, as can be inferred from Eqs.~\eqref{eq:fL} and~\eqref{eq:fR}.

Since the collider detectors do not distinguish the $d$-quark and $u$-quark, we need to identify $c_{W_\text{hel}}\longleftrightarrow -c_{W_\text{hel}}$. However, the quark emitted in the forward direction in the $W$ rest frame will be harder and more separated from the $b$-quark in the top rest frame. Similarly, the quark emitted in the backward direction in the $W$ frame will be softer and more aligned with the $b$-quark in the top rest frame. Therefore, the hard and soft quarks have a probability of being the $d$-quark given by 
\begin{eqnarray}
    p(d\rightarrow q_\text{hard}) &=& \frac{\rho(\abs{c_{W_\text{hel}}})}{\rho(\abs{c_{W_\text{hel}}}) + \rho(-\abs{c_{W_\text{hel}}})} \, ,\\
    p(d\rightarrow q_\text{soft}) &=& \frac{\rho(-\abs{c_{W_\text{hel}}})}{\rho(\abs{c_{W_\text{hel}}}) + \rho(-\abs{c_{W_\text{hel}}})} \,  .
\end{eqnarray}
    
The top spin axis will align with the direction defined by the weighted average of the hard and soft-quark directions~\cite{Tweedie:2014yda}, so the optimal direction is given by 
\begin{eqnarray}
    \vec{q}_\text{opt}(\abs{c_{W_\text{hel}}}) &=& p(d\rightarrow q_\text{hard}) \, \hat{q}_\text{hard} \nonumber  \\
    &&  + p(d\rightarrow q_\text{soft})\, \hat{q}_\text{soft}, 
    \label{eq:qopt}
\end{eqnarray}
where $\hat{q}_\text{hard}$ and $\hat{q}_\text{soft}$ denote the direction of the hard and soft quarks in the top rest frame, respectively. 
The spin analyzer power, $\beta_\text{opt}$, as a function of $\abs{c_{W_\text{hel}}}$ is equal to the length of $\vec{q}_\text{opt}(\abs{c_{W_{hel}}})$. Its integrated value is $\beta_\text{opt}\simeq 0.64$.

\subsection{Quantum tomography}

The top quark pair  $t\bar t$ forms a two-qubit system that can be represented by the spin density matrix
\begin{align}
    \rho=\frac{\mathbb{1}\otimes \mathbb{1}+\sum_i \left(B_i\sigma_i\otimes \mathbb{1} + \bar{B}_i \mathbb{1}\otimes\sigma_i\right)+\sum_{ij}C_{ij}\sigma_i\otimes\sigma_j}{4},
    \label{eq:density}
\end{align}
where $B_i$ and $\bar{B}_i$  are the spin polarizations and $C_{ij}$ are the spin correlations for the spin-$1/2$ particles. The spin density matrix can be simplified when applied to strong $t\bar t$ production at the LHC. Strong top pair production satisfies the P and CP symmetries, leading to $B_i=\bar{B}_i=0$ and $C_{ij}=C_{ji}$~\cite{Bernreuther:2015yna}. Electroweak corrections can change these relations; however, they have been shown to be subleading~\cite{Frederix:2021zsh}. In combination with these results, the $t\bar t$ density matrix can be further simplified in the helicity basis, where the only nonvanishing parameters are the diagonal entries $C_{ii}$ and one off-diagonal term $(C_{12}\simeq C_{21})$. 
 The other off-diagonal terms are 
generated by P-odd absorptive parts of the mixed QCD-weak corrections at one-loop, and are negligible \cite{Bernreuther:2015yna}. Henceforth, we shall assume these well-justified relations. In this study, the top quark spin is measured in the helicity basis $(\hat{k}, \hat{r},\hat{n})$~\cite{Bernreuther:2010ny}:
\begin{itemize}
    \item $\hat{k}$ is the direction of the top quark momentum in the top pair rest frame.
    \item $\hat{r}=\text{sign}(\cos\theta)(\hat{p}-\cos\theta\hat{k})/\sin\theta$, where the beam axis $\hat{p}=(0,0,1)$ and $\cos\theta=\hat{k} \cdot \hat{p}$.
    \item $\hat{n}=\hat{k}\times \hat{r}$.
\end{itemize}

It was shown that the entanglement and Bell inequalities violation criteria are written in simpler forms in the helicity basis~\cite{Aguilar-Saavedra:2022uye}. More concretely, for the $t\bar t$ system, a condition for the negativity of $\rho^{T_{2}}$ is equivalent to the following relation
\begin{align}
   \mathcal{E}\equiv |C_{kk}+C_{rr}|-C_{nn}-1>0. \label{eq:E}
\end{align}
The Peres-Horodecki criterion is equivalent to the positivity of $\mathcal{E}$. The $\mathcal{E}$ is also related to the concurrence $\mathcal{C}(\rho) = \frac{1}{2}\text{max}[\mathcal{E},0]$ with $C_{nn} \leq 0$ and therefore its magnitude can be used as a measure of entanglement \cite{Afik:2022kwm}.
Analogously, the CHSH inequalities can be expressed in terms of spin correlations 
\begin{align}
\label{eq:B1}    \mathcal{B}_1&\equiv |C_{rr}-C_{nn}|-\sqrt{2} > 0,\\
\label{eq:B2}    \mathcal{B}_2&\equiv |C_{kk}+C_{rr}|-\sqrt{2} > 0, \\
\label{eq:B3}    \mathcal{B}_3&\equiv |C_{kk}+C_{nn}|-\sqrt{2} > 0, \\
\label{eq:B4}    \mathcal{B}_4&\equiv |C_{rr}+C_{nn}|-\sqrt{2} > 0.
\end{align}
We focus on $\mathcal{B}_1$ and $\mathcal{B}_2$ which lead to stronger bounds among others, since $C_{nn} <0$ and the coefficients $C_{rr}$ and  $C_{kk}$ are mostly positive in the $(m_{t\bar t}, \cos\theta_{CM})$ plane, as we explicitly show in the following section. The coefficients $C_{ij}$ can be measured with the distributions from the top and anti-top quark decay products $a$ and $b$
\begin{align}
    \frac{1}{\sigma}\frac{d^2\sigma}{d\cos\theta^i_a d \cos\bar\theta^j_b}=\frac{1}{4}\left(1+\beta_a\beta_b C_{ij}\cos\theta^i_a \cos\bar\theta^j_b\right)\,,
\label{eq:dist}
\end{align}
where $\theta_a^i$ is the polar angle for the final state $a$ with respect to the $i$th axis in the top quark rest frame and $\bar\theta_b^j$ is the polar angle for the final state $b$ with respect to the $j$th axis in the anti-top rest frame. 
The $\beta_a$ and $\beta_b$ are the spin analyzing powers for the corresponding final states.
We can single out the $C_{ij}$ coefficients from Eq.~\eqref{eq:dist} as 
\begin{align}
C_{ij}=\frac{4}{\beta_a\beta_b}\frac{N(c_{\theta_a^i}c_{\bar\theta_b^j}>0)-N(c_{\theta_a^i}c_{\bar\theta_b^j}<0)}{N(c_{\theta_a^i}c_{\bar\theta_b^j}>0)+N(c_{\theta_a^i}c_{\bar\theta_b^j}<0)}\,,
\label{eq:cij}
\end{align}
where $c_{\theta}\equiv \cos\theta$,  $c_{\bar \theta}\equiv \cos\bar\theta$, and $N$ denotes the number of events.

\begin{figure*}[th!]
\centering
    \includegraphics[width=0.32\textwidth]{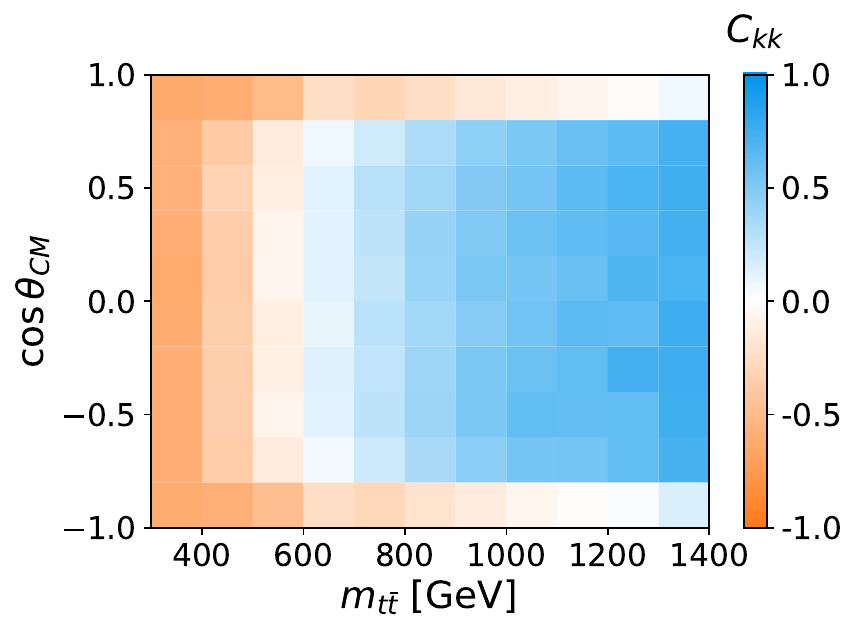}
    \includegraphics[width=0.32\textwidth]{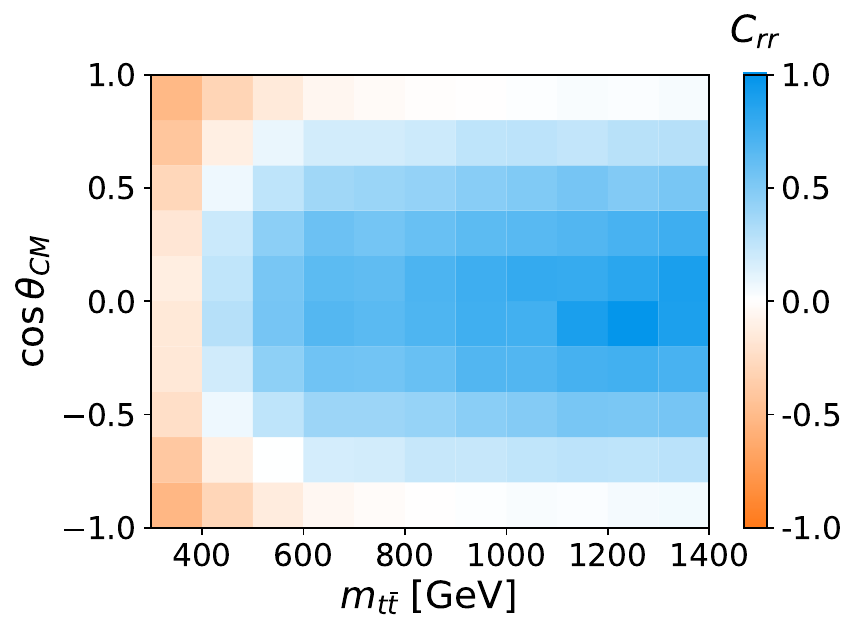}
    \includegraphics[width=0.32\textwidth]{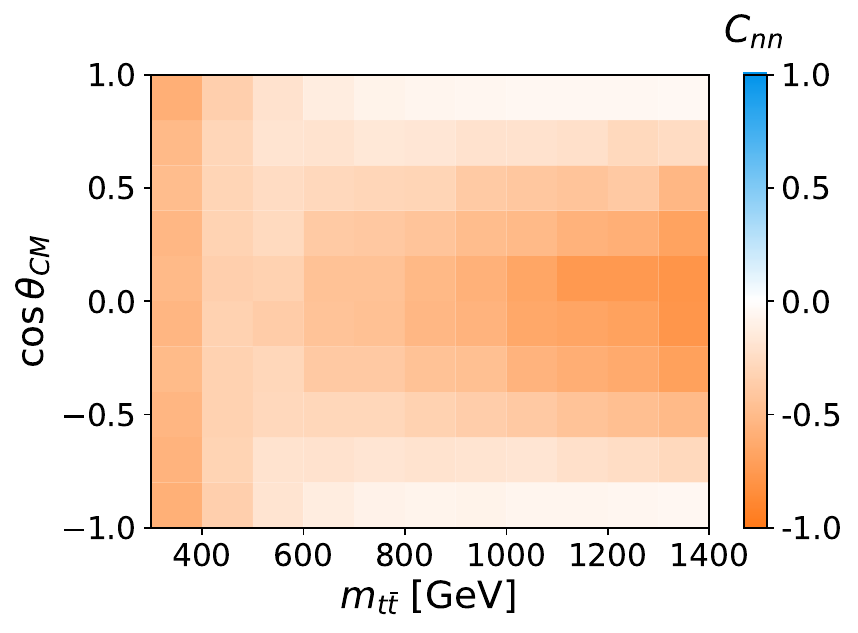}
     \vspace*{-0.3cm}
    \caption{Parton level distributions for the coefficients $C_{kk}$ (left), $C_{rr}$ (center), and $C_{nn}$ (right) in the helicity frame as a function of ($m_{t\bar{t}}, \cos\theta_\text{CM}$). The coefficients are obtained  with the top and anti-top quark decay products     \{$\ell^\pm,\vec{q}_\text{opt}$\}. The optimal hadronic direction $\vec{q}_\text{opt}$ is defined in Eq.~\eqref{eq:qopt}.
    \label{fig:parton_coeff} }
\end{figure*}

 \begin{figure*}
 \centering
     \includegraphics[width=0.32\textwidth]{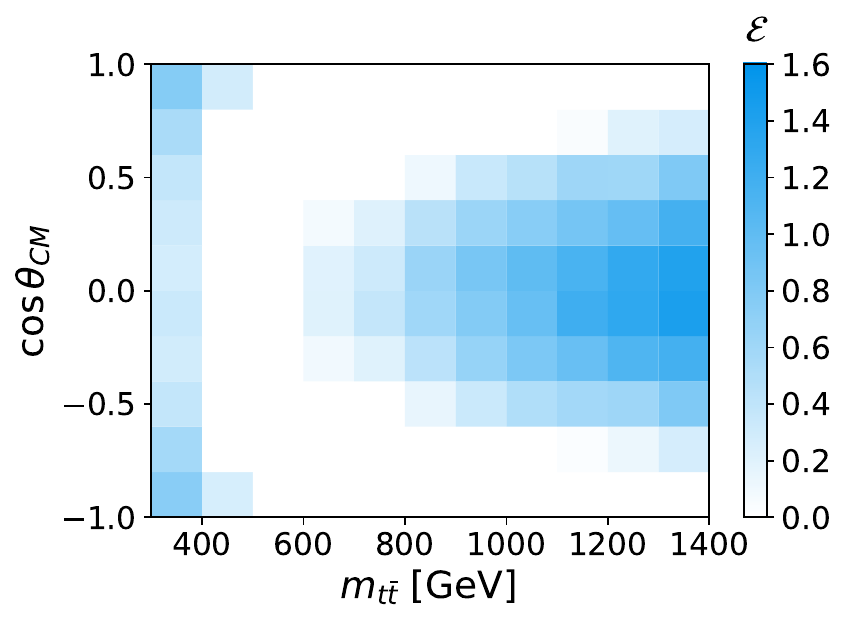}
     \includegraphics[width=0.32\textwidth]{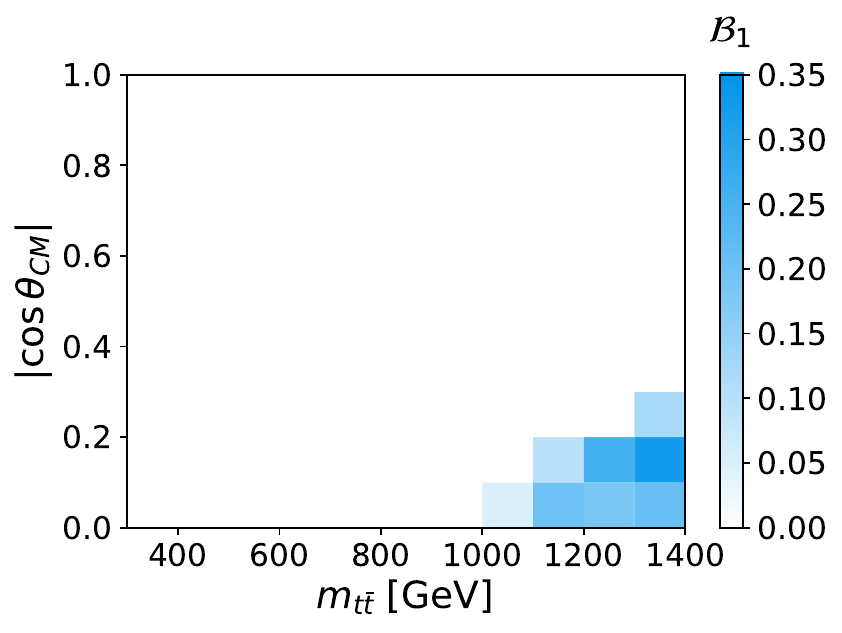}
     \includegraphics[width=0.32\textwidth]{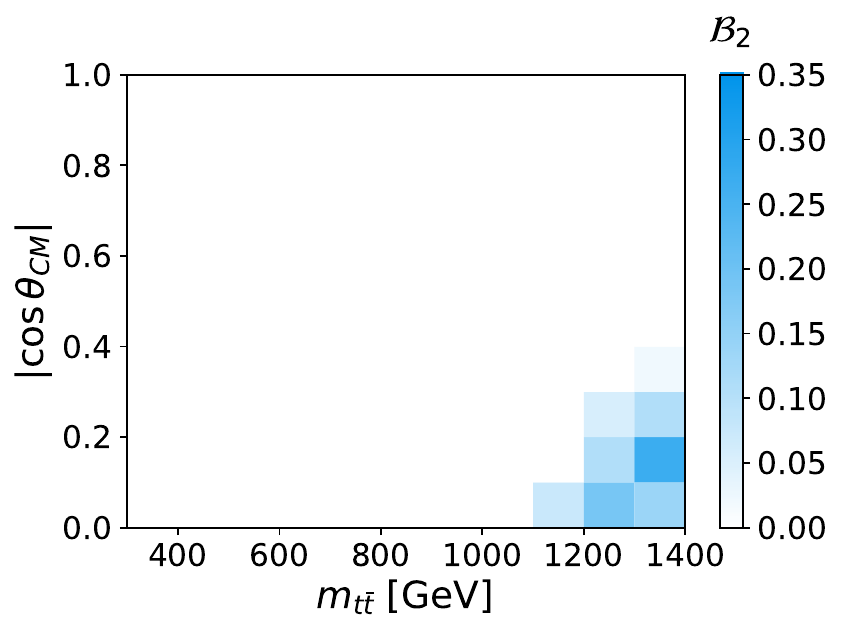}
     \vspace*{-0.3cm}
     \caption{Entanglement probe $\mathcal{E}$ (left) , CHSH violation indicators $\mathcal{B}_1$ (middle), and $\mathcal{B}_2$ (right) in the $(m_{t\bar{t}},\cos\theta_\text{CM})$ plane at parton level, using the charged lepton $\ell^\pm$ and the optimal hadronic direction $\vec{q}_\text{opt}$ defined in Eq.~\eqref{eq:qopt}.
     }
     \label{fig:parton_ent}
 \end{figure*}

\subsection{Spin correlations, entanglement, and Bell inequalities}
\label{sec:spincorrelation}

It is worth highlighting the distinction among the concepts of spin correlations, entanglement, and Bell inequalities. Spin correlations in top quark physics have been tested since the Tevatron era~\cite{PhysRevD.53.4886,Mahlon:2010gw,CDF:2010yag,D0:2011psp} and continue to be an important subject of study at the LHC~\cite{ATLAS:2019zrq,CMS:2019nrx}. They provide relevant probes for precision physics and beyond the Standard Model physics searches~\cite{Buckley:2015vsa,Buckley:2015ctj,Goncalves:2016qhh,Goncalves:2018agy,Goncalves:2021dcu,Barman:2021yfh}. The existence of spin correlations is ultimately a reflection of nonvanishing coefficients $C_{ij}$ in the spin density matrix, Eq.~\eqref{eq:density}. 

In contrast, the requirements for entanglement and violation of Bell inequalities further constrain the spin density matrix. In addition to requiring nonzero contributions $C_{ij}$, these conditions impose particular relations among the coefficients $C_{ij}$, as presented in Eq.~\eqref{eq:E}, and Eqs.~(\ref{eq:B1})-(\ref{eq:B4}). Regarding the comparison between the conditions for entanglement and violation of Bell inequalities, it is important to stress that entanglement is a necessary condition for the violation of Bell inequalities. However, the reverse is not true. The hierarchy of these conditions on the spin density matrix can be summarized as follows:
\begin{align}
    \text{Spin correl.}\supseteq \text{Entanglement} \supseteq \text{Bell inequalities violation}.
\end{align}

\section{Parton Level Distributions}
\label{sec:parton}

In this section, we analyze distributions for angular coefficients, entanglement indicator, and CHSH violation probe at parton level. The results obtained here will motivate our boosted top quark analysis presented in Sec.~\ref{sec:analysis}.

We start the parton level study generating a $pp\to t\bar t\to \ell^\pm \nu 2b 2j$ sample at the $\sqrt{s}=14$~TeV LHC, where  $\ell^\pm=e^\pm$ or $\mu^\pm$, using \texttt{M\textsc{ad}G\textsc{raph}5\_\textsc{a}MC@NLO}~\cite{Alwall:2014hca} at leading order within the standard model with \texttt{\textsc{nnpdf2.3qed}} for parton distribution function~\cite{Ball:2013hta}. The factorization and renormalization scales are set to $\mu_F=\mu_R=\left(\sqrt{m_t^2+p_{ T t}^2}+\sqrt{m_t^2+p_{T \bar{t}}^2} \right)/2$. 
We normalize the $t\bar t$ production cross section to $\sigma=985.7$~pb, as calculated with the Top++2.0 program to next-to-next-to-leading order in perturbative QCD, including soft-gluon resummation to next-to-next-to-leading-log order~\cite{Czakon:2011xx,Beneke:2011mq}. 
No event selections have been applied during the Monte Carlo (MC) generation.
Here, we present the kinematic regime with maximum sensitivity to these probes and show that the hadronic polarimeter defined in  Sec.~\ref{sec:pola} works as an efficient proxy for the $d$-quark.

\begin{figure*}[th!]
    \centering
    \includegraphics[width=0.33\textwidth]{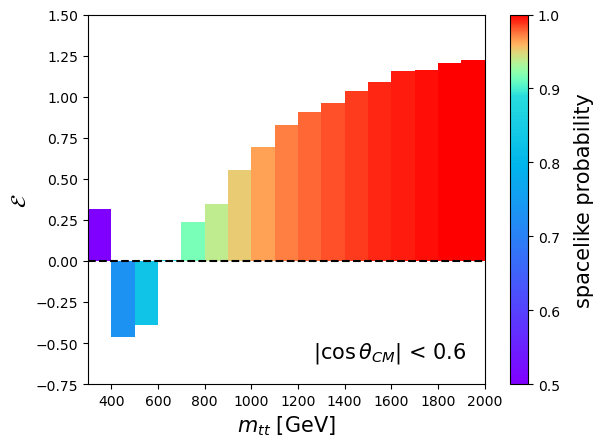} \hspace*{-0.2cm}
    \includegraphics[width=0.33\textwidth]{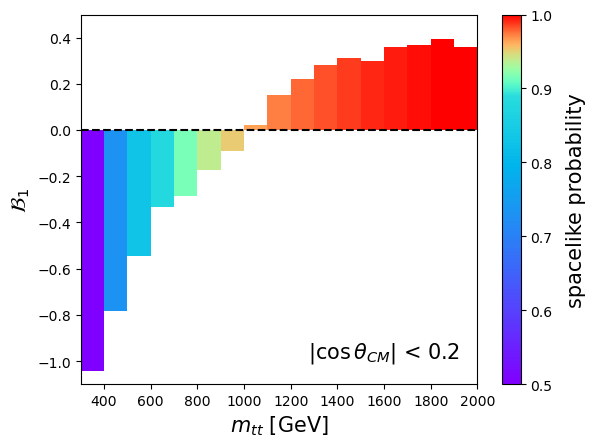}\hspace*{-0.2cm}
    \includegraphics[width=0.33\textwidth]{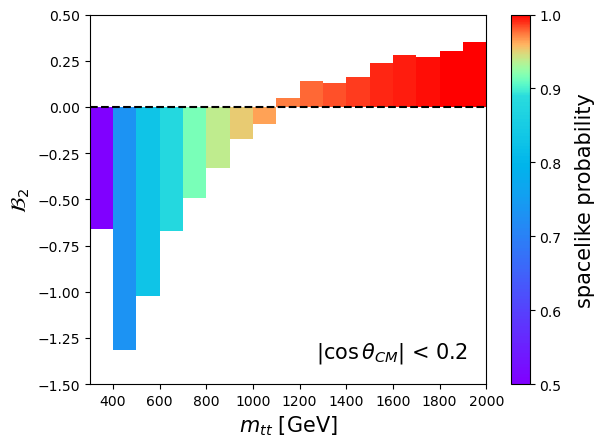}
    \caption{Entanglement ($\mathcal{E}$) and Bell inequalities measure ($\mathcal{B}_1, \mathcal{B}_2$) (at parton-level) as a function of $m_{tt}$ and spacelike separation probability.  The latter refers to  the fraction of $t$ and $\bar t$ decays that are spacelike-separated. We impose $|\cos\theta_\text{CM}|\leq 0.6$ for the left panel and $|\cos\theta_\text{CM}|\leq 0.2$ for the middle and right panels.} 
    \label{fig:E_B1_B2}
\end{figure*}

In Fig.~\ref{fig:parton_coeff}, we present the coefficients $C_{kk}$ (left), $C_{rr}$ (middle), and $C_{nn}$ (right)  in the helicity frame as a function of $(m_{t\bar t},\cos\theta_\text{CM})$, where $\theta_\text{CM}$ is the top quark scattering angle in the $t\bar{t}$ center-of-mass frame. The coefficients are derived with the top and anti-top decay products $\ell^\pm$ and $\vec{q}_\text{opt}$, where $\vec{q}_\text{opt}$ is the optimal hadronic direction defined in Eq.~\eqref{eq:qopt}. The coefficients are obtained with Eq.~\eqref{eq:cij}, assuming perfect reconstruction of the neutrino momentum. The coefficients for the spin density matrix display large dependencies with the kinematic regime. In the boosted region (large $m_{t\bar t}$) for a small $\cos\theta_{CM}$, $C_{kk}$ and $C_{rr}$ approach $1$ (represented by dark blue), while $C_{nn}$ approaches $-1$ (represented by dark orange), which maximizes  $\mathcal{E}$, $\mathcal{B}_1$, and $\mathcal{B}_2$ [see~\cref{eq:E,eq:B1,eq:B2,eq:B3,eq:B4}].

The numerical variation for the coefficients results in sizable kinematic correlations in the entanglement indicator $\mathcal{E}$ and CHSH probes $\mathcal{B}_{1,2}$, as illustrated in Fig.~\ref{fig:parton_ent}.
The blue shades represent the region of $(m_{t\bar t},\cos\theta_\text{CM})$ space where two top quarks are entangled ($\mathcal{E} > 0$) and violate the CHSH inequalities ($\mathcal{B}_{1,2} > 0$). Remarkably, there are two kinematic regions that display large values of $\mathcal{E}>0$, namely the $t\bar t$ production threshold  and high transverse momentum  ($m_{tt}\gg 2 m_t$ and $\theta_\text{CM}\sim \pi/2$). In contrast, CHSH inequalities are only violated in the high transverse momentum configuration.

The study of the boosted top quark regime for the semileptonic top pair process $pp\to t\bar t\to \ell \nu 2b 2j$ is motivated by three key reasons.  First, the semileptonic top pair has an event rate approximately $6$ times higher than that of the dileptonic process, making it a more effective probe for the high-energy regime. Second, in the high-energy configuration, boosted techniques can be employed to tag the hadronic top and efficiently identify the optimal hadronic direction, as described in the following section. Third, the two top quarks are expected to be causally disconnected from each other in the boosted regime~\cite{Severi:2021cnj}. 
Figure~\ref{fig:E_B1_B2} shows the causal disconnection, where the histogram heights represent the respective $(\mathcal{E}, \mathcal{B}_1, \mathcal{B}_2 )$ values as a function of the $t\bar t$ invariant mass $m_{t\bar t}$. The colors are mapped by the spacelike probability, which shows the fraction of $t$ and $\bar t$ decays that are spacelike separated, as a function of $m_{t\bar t}$. See Appendix~\ref{app:spacelike} for more details.  At the top quark pair threshold, a large fraction of the produced $t\bar{t}$ pairs are timelike separated; however, in the high-energy regime, spacelike separated top quark pairs are prevalent. For example, for $m_{t\bar t}>1$~TeV, approximately 95\% of the events display spacelike separated top quark pairs.
 
\section{Analysis}
\label{sec:analysis}

\begin{figure*}
    \centering
    \includegraphics[width=0.35\textwidth]{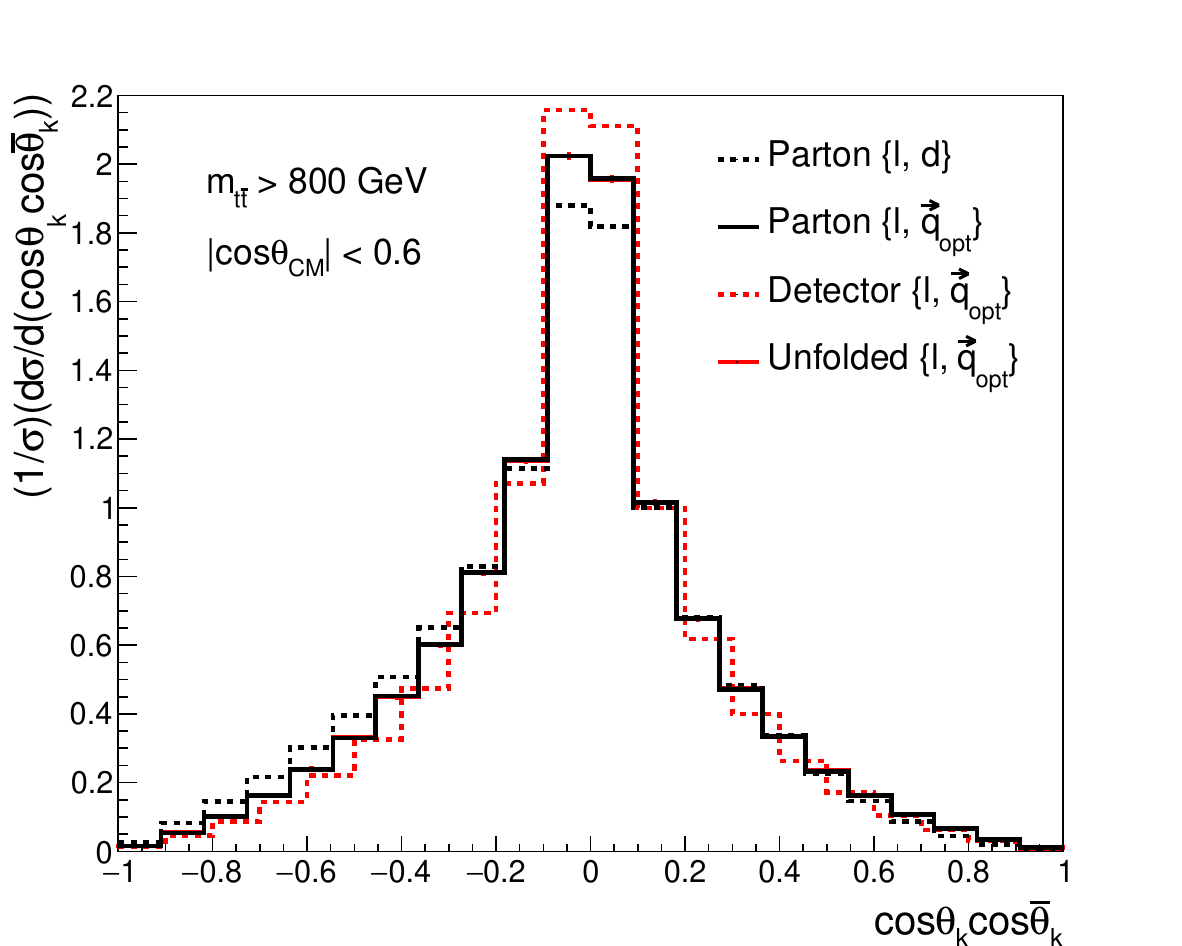} \hspace*{-0.7cm}
    \includegraphics[width=0.35\textwidth]{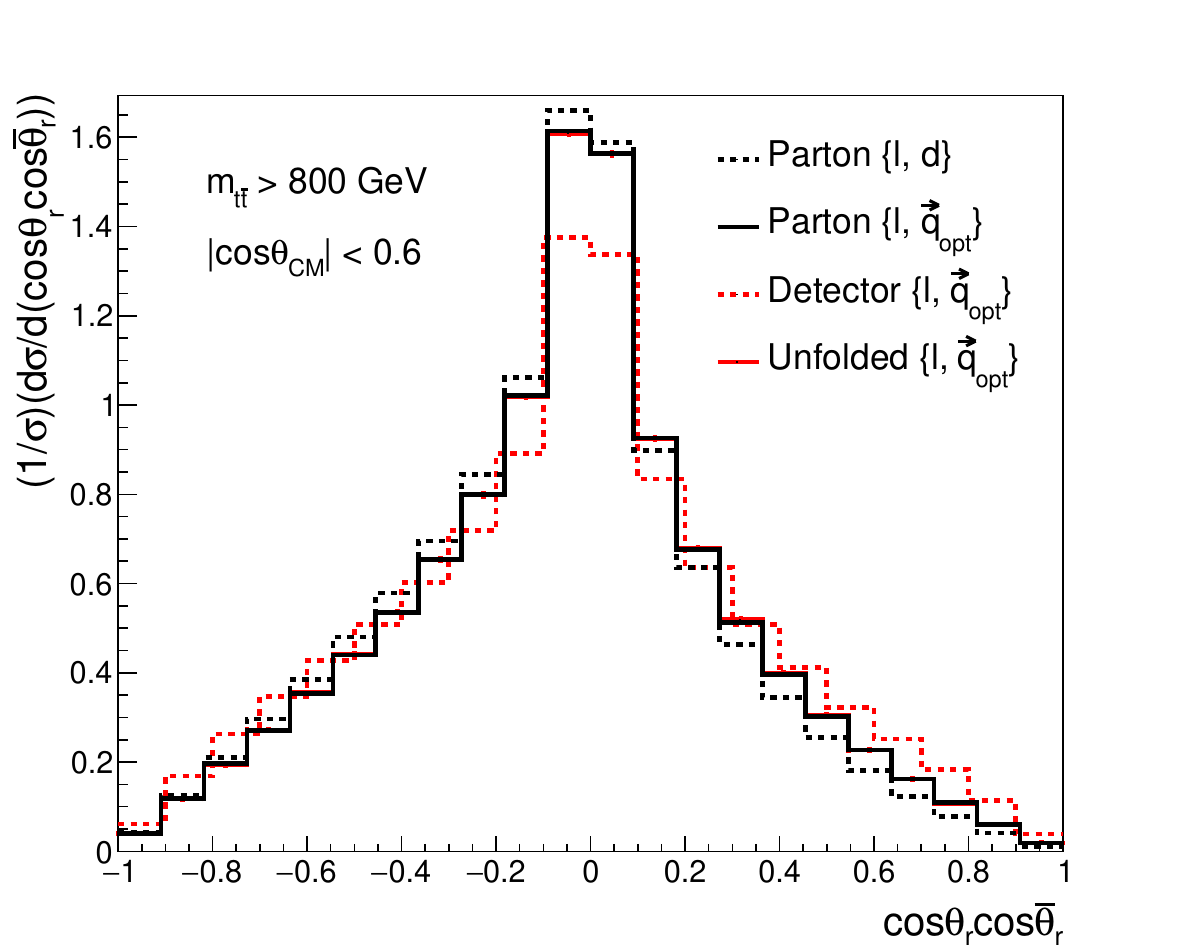}\hspace*{-0.7cm}
    \includegraphics[width=0.35\textwidth]{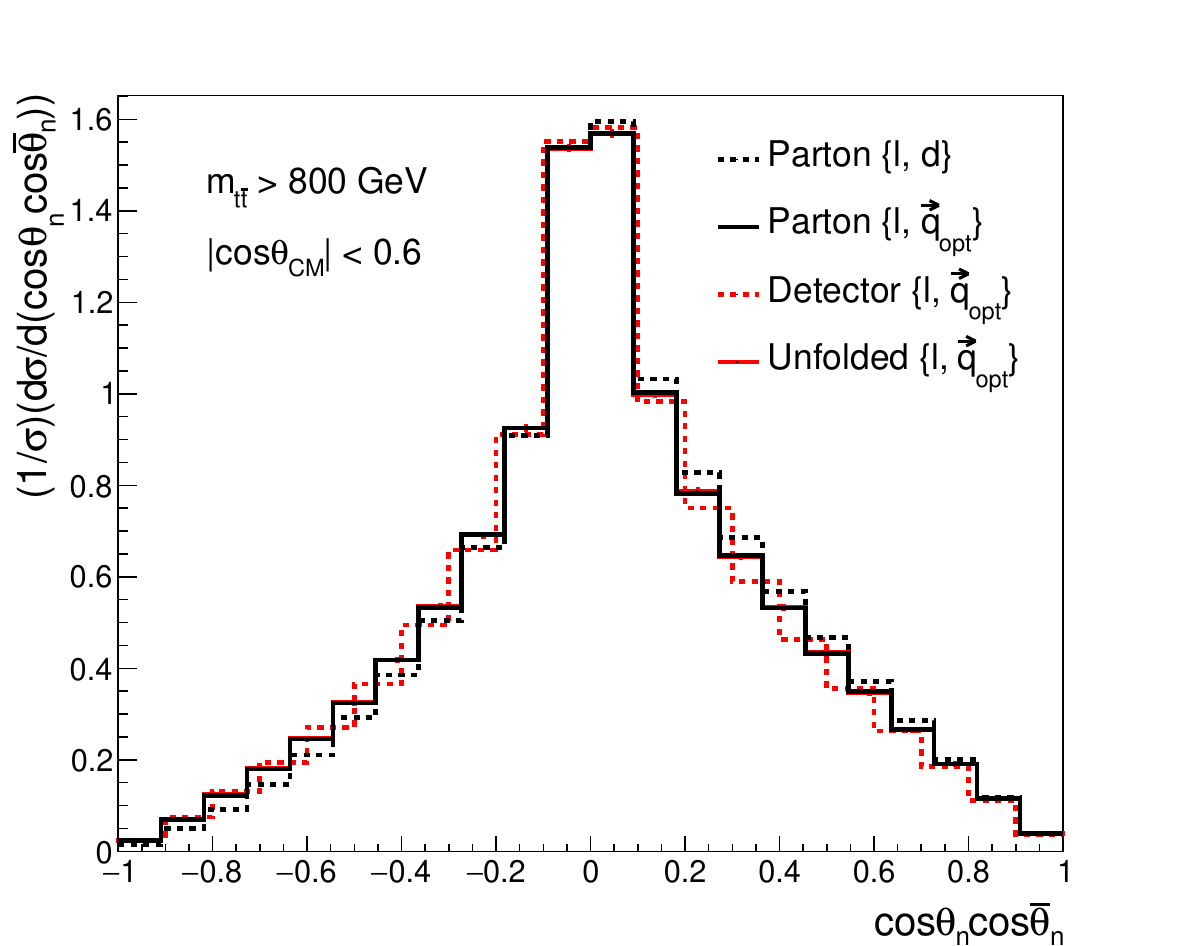} 
    \caption{Differential angular distributions with respect to $\cos\theta^i_\ell \cos\bar \theta^i_{q_\text{opt}}$ for $i=k$ (left), $i=r$ (middle) and $i=n$ (right). We display the parton level correlations between the charged lepton and down-type quark (black dashed), and charge lepton with optimal direction (black solid). The detector level (red dashed)  and unfolded (red solid) results for the correlations between charged lepton and optimal hadronic direction are also presented. We apply the selections $m_{t\bar{t}}\geq 800$ GeV and $|\cos\theta_\text{CM}|\leq 0.6$.
    \label{fig:cosicosj} }
\end{figure*}

In this section, we derive the sensitivity for the entanglement probe $\mathcal{E}$ and CHSH violation indicators $\mathcal{B}_{1,2}$ at the HL-LHC. We focus on the boosted regime, using the semileptonic top pair final state. 
Parton-level events are further processed, accounting for parton shower and hadronization effects using~\texttt{\textsc{pythia8}}~\cite{Sjostrand:2007gs}. 
We simulate the detector effects for all jets according to ATLAS parametrization~\cite{TheATLAScollaboration:2013sgb}, and the missing transverse momentum $\mptvec$ by Gaussian smearing with resolution $\sigma = -0.07+2.92/|\mptvec|$~\cite{deFavereau:2013fsa}. We begin our analysis by requiring one isolated charged lepton $\ell^\pm=\{e^\pm, \mu^\pm\}$ with $p_{T\ell}>10$~GeV and $|\eta_\ell|<3$.
We then impose the missing transverse momentum cut, $\met = |\mptvec| > 30$~GeV and the minimum invariant mass $m_{t \bar t} > 600$ GeV. The goal here is to find four momenta of the hadronic and leptonic top quarks and reconstruct the spin correlation matrix $C_{ij}$, whose combinations would determine the entanglement probe and the Bell inequalities measurements. 

For the hadronic part of the event, instead of proceeding with the traditional resolved semileptonic $t\bar t$ analysis, we take advantage of the high top tagging efficiency in the boosted regime. We start by reconstructing jets with the Cambridge-Aachen algorithm 
with $R=1.5$~\cite{Cacciari:2011ma}, demanding at least one fat jet with $p_{TJ}>150$~GeV and $|\eta_J|<3$.
We require that one of the fat jets be tagged with the \texttt{HEPT\textsc{opTagger}}~\cite{Plehn:2010st}. To guarantee a robust reconstruction, the tagged top jet should satisfy the top mass constraint 130~GeV$<m_\text{top-jet}<215$~GeV. 
The high transverse momentum results in a large top tagging efficiency and a small fake rate for this algorithm. Furthermore, we observe that the boosted top quark regime dovetails nicely with both entanglement  and CHSH probes, respectively $\mathcal{E}$ and $\mathcal{B}_{1,2}$, which are enhanced at high-energy scales and display spacelike separated top pairs, as shown in Figs.~\ref{fig:parton_ent} and \ref{fig:E_B1_B2}. 

After tagging the hadronic top, we require one of its subjets to be $b$-tagged with efficiency $\epsilon_b=70\%$ and rejection factor of 300 (7) for light jets (charm-quark jets)~\cite{ATLAS:2022mlu}.  We tailored the \texttt{\textsc{HEPTopTagger}}'s $b$-subjet identification process by incorporating machine learning algorithms. Although the standard  \texttt{\textsc{HEPTopTagger}} approach mostly depends on the mass relationships among the three subjets $(m_{12},m_{13},m_{23})$~\cite{Plehn:2010st}, our custom version for the $b$-jet identification employs additional machine learning techniques.
Once three subjets within the top jet are identified using the top tagging algorithm, we run the Lorentz Boost Network (LBN)~\cite{Erdmann:2018shi} to determine which one of the three subjects is the $b$-jet  candidate. The network takes the four-momenta of the three subjets as inputs ($p_T$-ordered), and outputs a categorical label (1 of 3). We used three hidden layers, with the LBN-specific hyperparameter ``M'' (the number of hypothetical particles or the number of rest frames) set to $12$. The network forms $12$ linear combinations of input particles and boosts them to different frames: then the features of the boosted particles are fed into a simple deep neural network (DNN). The set of features of the boosted particles is selected to be (energy, transverse momentum, $\eta,~\phi$, and mass) of each particle, as well as the cosine of the angle between each pair of particles. The network is trained on subjets identified by the standard~\texttt{\textsc{HEPTopTagger}} algorithm. The efficiency and purity of the LBN with $b$-tagging criteria (checking the $B$-hadrons inside the subjet) is 75\% and 96\%, respectively, while they are 68\% and 94\% when using the default \texttt{\textsc{HEPTopTagger}}. Hence, the main advantage of using \texttt{\textsc{HEPTopTagger}}+LBN would be the higher efficiency when compared to the standard top tagging algorithm.

After identifying the $b$-tagged subjet candidate, one can select a proxy for the down-type jet candidate from the remaining two subjets. One common strategy is to choose the softer of the two light subjets in the rest frame of the top jet. This initial choice provides a spin analyzing power of about $\beta_\text{soft}\sim 0.5$~\cite{Jezabek:1994qs,Barman:2021yfh}. However, the spin analyzing power can be improved by selecting an optimal hadronic polarimeter, which results in $\beta_\text{opt}\approx 0.64$~\cite{Tweedie:2014yda}. We follow the latter approach. See Sec.~\ref{sec:pola} for further details. 
 
Since we have only one heavy hadronic particle in our signal event, namely the hadronic top quark, we study the remaining event reconstruction using a smaller jet  radius. This suppresses the underlying event contamination. Thus, we remove all the hadronic activity associated with the successfully top-tagged jet and recluster the remaining hadronic activity by applying the anti-$k_T$ jet algorithm with radius $R=0.4$. Jets are defined with $p_{Tj}>30$~GeV and $|\eta_j|<3$. 
We demand a $b$-tagged jet in the leptonic top decay, assuming a $b$-tagging efficiency of 85\% and 1\% mistag rate for light jets. This large $b$-tagging performance is in agreement with the HL-LHC projections performed by ATLAS~\cite{CERN-LHCC-2017-005}, which account for the new central tracking systems that will be in operation at the HL-LHC.
Table \ref{table:cutflow} summarizes the reconstruction procedure that we have described so far, and the corresponding cross sections.
\begin{table*}[t]
\renewcommand\arraystretch{1.5}
\begin{tabular}{|c||c|}
\hline
Reconstruction procedure & Cross sections \\ \hline \hline
    NNLO cross section for $pp \to t \bar t$ at 14 TeV                      &  ~~$\sigma_\text{NNLO}=985.7$~pb~~ \\
\hline
    Minimum invariant mass $m_{t \bar t} > 600$ GeV  &  $20.0$ pb \\
\hline
One isolated lepton $\ell^\pm=\{e^\pm, \mu^\pm\}$ with $p_{T\ell}>10$~GeV and $|\eta_\ell|<3$, and  & \\
 missing transverse momentum $\met = |\mptvec| > 30$~GeV                                     & $34.5$ pb \\
\hline
At least one fat jet $J$ with $R=1.5$, $p_{TJ}>150$~GeV and $|\eta_J|<3$, &  \\
one of the fat jets has to be tagged with the \texttt{\textsc{HEPTopTagger}}, & $3.52$ pb \\
~~and one of the subjets has to be $b$-tagged ($\epsilon_{b\to b}=0.7$, $\epsilon_{j \to b}=1/300$, $\epsilon_{c \to b}=1/7$)~~& \\
\hline
~~At least one $b$-tagged jet with $R=0.4$, $p_{Tj}>30$~GeV, and $|\eta_j|<3$ ($\epsilon_{b\to b}=0.85$, $\epsilon_{j\to b}=0.01$) ~~ & $2.76$ pb \\
\hline
$m_{t\bar t} > 800$ GeV and $|\cos\theta_{\rm CM}| < 0.6$ for entanglement analysis & 382.8 fb \\
\hline
$m_{t\bar t} > 1.3$ TeV and $|\cos\theta_{\rm CM}| < 0.2$ for Bell's inequality analysis & 8.25 fb\\
\hline
\end{tabular}
\caption{Cumulative cut-flow table showing cross section for $t\bar t$ production in the semileptonic channel.\label{table:cutflow}}
\end{table*}

The dominant backgrounds are $W + {\rm jets}$ and single top production $tW$   followed by other subleading contributions from $t\bar t V$, $t\bar t h$, $Z + {\rm jets}$, and diboson production. 
With a relatively high invariant mass cut and the signal selection criteria, ATLAS Collaboration finds that these backgrounds are as small as $\sim 4\%$ of $t\bar t$ production~\cite{ATLAS:2022xfj}. With hadronic top tagging, the background will be even further suppressed in our analysis framework, resulting in approximately 2\% of the signal rate. This strategy, in particular, effectively mitigates the dominant background contribution from $W + {\rm jets}$. Refer to Appendix~\ref{app:backgrounds} for more comprehensive information regarding the background estimation.

A simple approach to find the momentum of the leptonic top quark would be via analytic reconstruction. One can use the on-shell condition of top quark or $W$ boson to fix the $z$-component of the neutrino momentum. However, considering the finite width effects and the poor detector resolution for the missing transverse momentum, it frequently leads to a complex solution. To overcome these problems, one can scan over the three momentum of the neutrino until the on-shell conditions are met. Alternatively, one can attempt to find all three components of the missing neutrino via 
\begin{align}
    \chi^2 = \dfrac{(m_t-m_{b\ell\nu})^2}{\sigma_t^2}+\dfrac{(m_W-m_{\ell\nu})^2}{\sigma_W^2}+\dfrac{(\mptvec-\vec p_{\nu T})^2}{\sigma_{MET}^2},
\end{align}
where $\sigma_t$, $\sigma_W$, and $\sigma_{MET}$ are the mass resolution of top quark, $W$ boson, and the uncertainty in the missing transverse momentum measurement, respectively. While the mass fits are very good, as required by construction, we find that the shape of angular distributions with the top quark momentum obtained via $\chi^2$ can significantly differ from the expected shape of parton-level distributions.

To overcome this problem, in the present study, we used the LBN with the following input features: the four-momenta of the lepton and $b$-jet associated with the leptonic top, as well as the $x$ and $y$ components of $\mptvec$ as an artificial four-vector, with the energy and $z$ component set to zero. We employed the same LBN setup as in the hadronic top case, with the output layer consisting of three dimensions, corresponding to the three-momentum component of the neutrino. 
 
 To train the network, we developed a custom loss function defined as $L = \text{Mean squared loss}+\lambda_1(m_{b\ell\nu}-m_t)^2+\lambda_2(m_{\ell\nu}-m_W)^2$, where $m_{b\ell\nu} $ and $ m_{\ell\nu}$ represent the invariant masses of the reconstructed top quark and  $W$ boson, and $m_t$, $m_W$ are their true masses. The hyperparameters $\lambda_1$ and $\lambda_2$ were tuned to 0.8 and 0.4, respectively. We find that the LBN method outperforms the $\chi^2$ approach in terms of correct angular distributions. 

Once we find the momenta of the hadronic and leptonic top quarks, we compute the double angular distributions ${d^2\sigma}/{d\cos\theta^i_\ell d \cos\bar\theta^j_{\rm q_\text{opt}}}$ in the helicity basis as in Eq.~(\ref{eq:dist}).
Figure~\ref{fig:cosicosj} shows the differential angular distributions with respect to $\cos\theta^i_\ell  \cos\bar\theta^i_{q_\text{opt}}$ for $i=k$ (left), $i=n$ (middle), and $i=r$ (right). 
Parton-level distributions with the charged lepton and down-type quark (dashed lines) and optimal hadronic direction (solid lines) are shown in black, while detector-level distributions with the optimal direction (dashed lines) and unfolded distributions (solid lines) are shown in red, respectively. We observe that the profiles of parton-level distributions are very well retained at each stage of the analysis.

As shown in Fig.~\ref{fig:cosicosj}, the overall angular distributions do not change significantly in these different stages of the simulation. However, the correlation coefficients are very sensitive to small effects in the angular distributions (due to cuts, detector effects, imprecise estimation of the optimal hadronic direction, etc.), affecting the entanglement and Bell inequalities estimations. To reach a more robust quantitative conclusion,  we perform unfolding with the \textsc{TSVDUnfold} package~\cite{H_cker_1996}. It uses singular value decomposition (SVD) of the response matrix. 
Details of the unfolding algorithm and adopted parameters are described in Appendix~\ref{app:unfold} (see Ref.~\cite{Tweedie:2014yda} for the stability of the optimal direction).

\begin{figure}[t]
    \centering
    \includegraphics[width=0.45\textwidth]{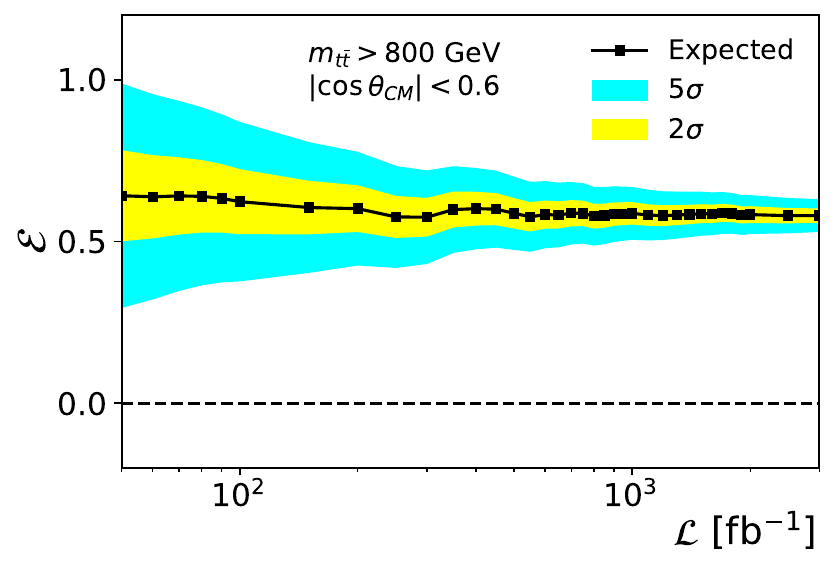}
    \caption{Entanglement indicator as a function of the luminosity. 
    The yellow and blue areas represent the regions for $\mathcal{E}\pm 2\sigma$ and $\mathcal{E}\pm 5\sigma$, respectively.}
    \label{fig:Ent_unf}
\end{figure}

\begin{figure*}[th!]
    \centering
    \includegraphics[width=0.45\textwidth]{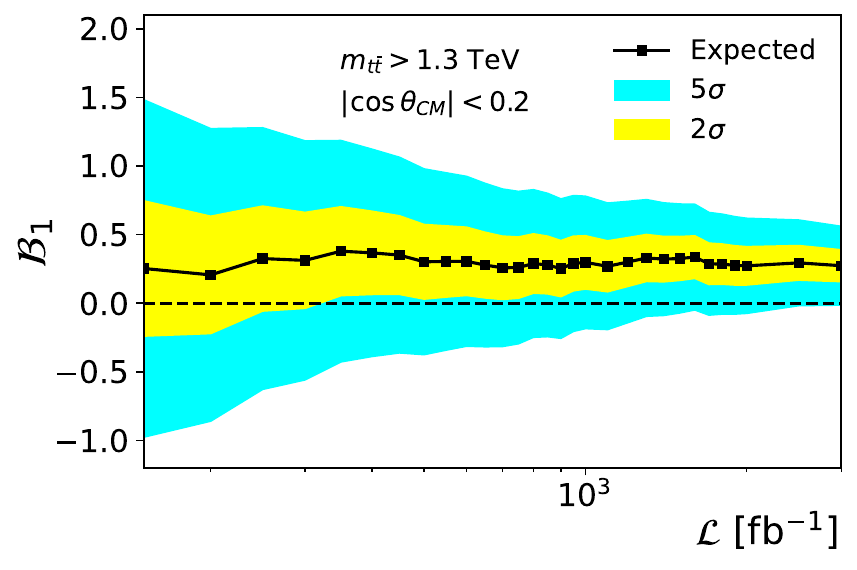}
    \hspace*{0.3cm}
    \includegraphics[width=0.45\textwidth]{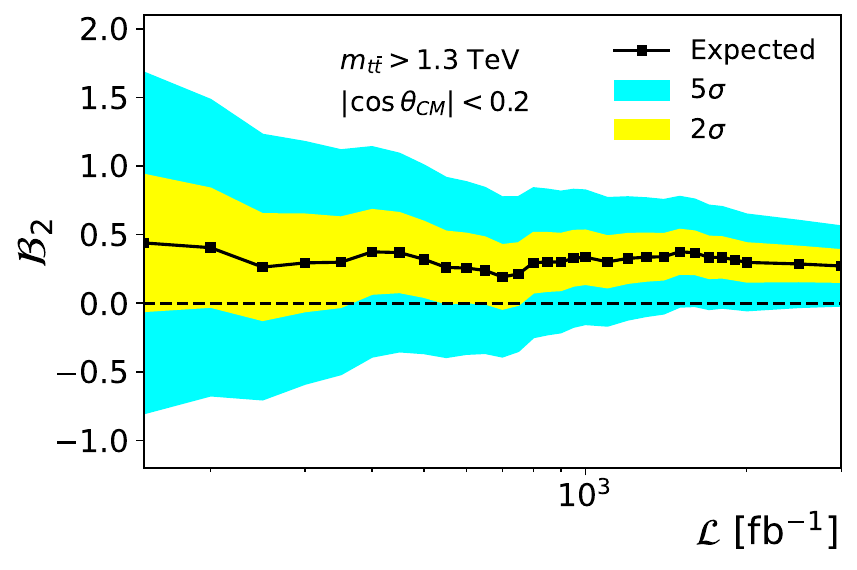}
    \caption{Bell indicators $\mathcal{B}_{1}$ (left panel) and $\mathcal{B}_{2}$ (right panel) 
    as a function of the luminosity. 
    The yellow and blue areas represent the regions for $\mathcal{B}_{i}\pm 2\sigma$ and $\mathcal{B}_{i}\pm 5\sigma$ ($i=1,2$), respectively.}
    \label{fig:Bells_unf_lumi}
\end{figure*}

\section{Results} 
To maximize the entanglement probe and CHSH violation indicator, we focus on the highly relativistic regime of ($m_{t\bar t}$, $\cos\theta_\text{CM}$) instead of the entire phase space. 
In the case of entanglement, we choose the region defined by $m_{t\bar{t}}\geq 800$ GeV and $|\cos\theta_\text{CM}|\leq 0.6$, which leads to the cross section of $382.8$~fb. We use approximately $7.8$~M events for training (corresponding to $20$~ab$^{-1}$ luminosity). 
The unfolded result for the entanglement probe $\mathcal{E}$ is shown in Fig.~\ref{fig:Ent_unf} as a function of the luminosity. 
We have randomly selected the events from the test dataset corresponding to the adopted 
luminosity. The yellow and blue regions represent the regions for $\mathcal{E}\pm 2\sigma$ and $\mathcal{E}\pm 5\sigma$, respectively, where $\sigma$ is the error estimated using \textsc{TSVDUnfold}~\cite{H_cker_1996}.
Remarkably, Fig.~\ref{fig:Ent_unf} indicates that ATLAS and CMS Collaborations have already accumulated sufficient data to observe entanglement between two top quarks, using the semileptonic top pair final state in the boosted regime.

For the Bell inequalities, we impose a more stringent phase space restriction, $m_{t\bar{t}}\geq 1.3$ TeV and $|\cos\theta_\text{CM}|\leq 0.2$, reducing the cross section to $8.25$~fb.\footnote{Refer to Appendix~\ref{app:loopholes} for a discussion on  loopholes associated with the measurement of Bell inequalities at LHC.} 
We randomly select the test data corresponding to $3$ ab$^{-1}$ and use the remaining data ($288$k events, corresponding to $35$~ab$^{-1}$ luminosity) for training.
Figure~\ref{fig:Bells_unf_lumi} displays the unfolded CHSH violation indicators, $\mathcal{B}_1$ (left) and $\mathcal{B}_2$ (right), as functions of luminosity.
Figure~\ref{fig:Bells_unf_lumi} indicates that the 5$\sigma$ observation of the CHSH violation is very promising with the full HL-LHC luminosity, especially when results from CMS and ATLAS can be combined.

\section{Summary}
\label{sec:summary}
\begin{table}
    \centering
    \renewcommand\arraystretch{1.5}
    \begin{tabular}{|c || c | c | c|}
    \hline
    ~Indicator~ & ~\thead{Parton-level}~ & ~Unfolded ~ & ~\thead{Significance \\ $(\mathcal{L}=3$ ab$^{-1}$)}~ \\   \hline 
      $\mathcal{B}_{1}$ & $0.267\pm 0.023$ & $0.274\pm 0.057$  & $4.8$ \\ \hline 
       $\mathcal{B}_{2}$ & $0.204\pm 0.023$ & $0.272\pm 0.058$ & $4.7$ \\ \hline 
    \end{tabular}
    \caption{Parton-level and unfolded values of $\mathcal{B}_{1}$ and $\mathcal{B}_{2}$ at the HL-LHC with $\mathcal{L}=3$ ab$^{-1}$. The parton-level uncertainty only accounts for the Monte Carlo error and is derived to obtain a robust benchmark.
    }
    \label{tab:my_label}
\end{table}

We performed a comprehensive analysis to investigate the feasibility of detecting quantum entanglement and the violation of Bell inequalities in top quark pair production at the LHC, with a focus on the semileptonic top pair final state. This study has significant implications, as it offers an excellent opportunity to test these quantum correlations at high-energy scales.

Both entanglement and violation of Bell inequalities are shown to be favored in the large invariant mass of the $t\bar{t}$ system. Boosted top tagging is very well motivated for such a kinematic regime. Using the semileptonic top pair final state, we take advantage of its $6$ times higher event rate compared to the dileptonic top pair final state, explored in previous studies~\cite{Afik:2020onf,Fabbrichesi:2021npl,Severi:2021cnj,Aguilar-Saavedra:2022uye}.

We showed that the final state with lepton and  optimal hadronic direction, in the semileptonic channel, can effectively probe the top quark pair spin density matrix, prompting access to the entanglement probe and the CHSH violation indicator. For more realistic simulation, we have included parton-shower, hadronization and detector effects. We have shown that \texttt{\textsc{HEPTopTagger}} and Lorentz Boost Network provide excellent reconstruction of the hadronic and leptonic top quarks.

Our final results indicate that the detection of entanglement for the semileptonic top pairs is straightforward, in agreement with existing studies in the literature that use dileptonic tops.\footnote{Following the submission of our work to arXiv, ATLAS reported the detection of quantum entanglement with dileptonic tops, focusing on the threshold regime~\cite{ATLAS:2023fsd}. The experimental study of the semileptonic boosted tops, initially proposed in the present article, is still pending as of the submission of this paper.}  ATLAS and CMS Collaborations should be able to observe entanglement in semileptonic tops with their current datasets. However, probing the violation of Bell inequalities proves more challenging, given its more restrictive nature compared to entanglement. Nevertheless, we establish the feasibility of such an observation at a significance level exceeding 4$\sigma$ at the HL-LHC (refer to Table~\ref{tab:my_label}). These results can be  statistically boosted even further by a combination of ATLAS and CMS datasets and accounting as well for the dileptonic top quark pair final state.

\section*{ACKNOWLEDGMENTS}
\label{sec:acknowledgements}

The authors thank Alexander Khanov, Mayukh Lahiri, Myeonghun Park, Joonwoo Bae, and John Ralston for useful discussions. D.G. and A.N. thank the U.S.~Department of Energy for the financial support, under Grant No. DE-SC 0016013. 
Z.D. is supported in part by the U.S. Department of Energy under Grant No. DE-SC0024407. K.K. is supported in part by the U.S. DOE under Award No. DE-SC0024407. This investigation was supported in part by the University of Kansas General Research Fund allocation 2151077. Some computing for this project was performed at the High Performance Computing Center at Oklahoma State University, supported in part through the National Science Foundation Grant No. OAC-1531128.

\quad

\appendix

\section{BACKGROUNDS}
\label{app:backgrounds}
Table~\ref{table:backgrounds} displays the cross sections for both signal and background after the complete reconstruction of the hadronic and leptonic top quarks, as described in Sec.~\ref{sec:analysis}. The cross sections are obtained in two stages: after additional cuts used for entanglement study ($m_{t\bar t} > 800$ GeV and 
$|\cos\theta_{\rm CM}| < 0.6$) in the first row, and Bell inequalities ($m_{t\bar t} > 1.3$ TeV and $|\cos\theta_{\rm CM}| < 0.2$) in the second row. 
In both cases, backgrounds display subleading effects, resulting in approximately $2\%$ of the signal rate. Notably, the imposition of the hadronic top tagging and the requirement of two $b$-tags play crucial roles in suppressing the $W+$jets and $tW$ backgrounds.
\begin{table}[h!]
\centering
\renewcommand\arraystretch{1.5}
\begin{tabular}{| c || c | c | c | c |} 
 \hline
   Cuts & ~~Signal~~ & ~~~$t\bar t V$~~~ & ~~~~$tW$~~~~ & ~$W$+jets~ \\ [0.5ex] 
 \hline\hline
 Entanglement & 382.8  & 0.74 & 7.27 & 1.03 \\ 
 \hline
 ~Bell inequalities~ & 8.25 & 0.02 & 0.14 & 0.02 \\ [1ex] 
 \hline
\end{tabular}
\caption{Cross sections for signal and backgrounds in $fb$ at the HL-LHC. $V$ includes $Z$, $W^\pm$ and $h$. Entanglement and Bell inequalities refer to the kinematic selections imposed on the respective analyses.
\label{table:backgrounds}}
\end{table}

\begin{figure*}[thb!]
    \centering
    \includegraphics[width=.35\textwidth]{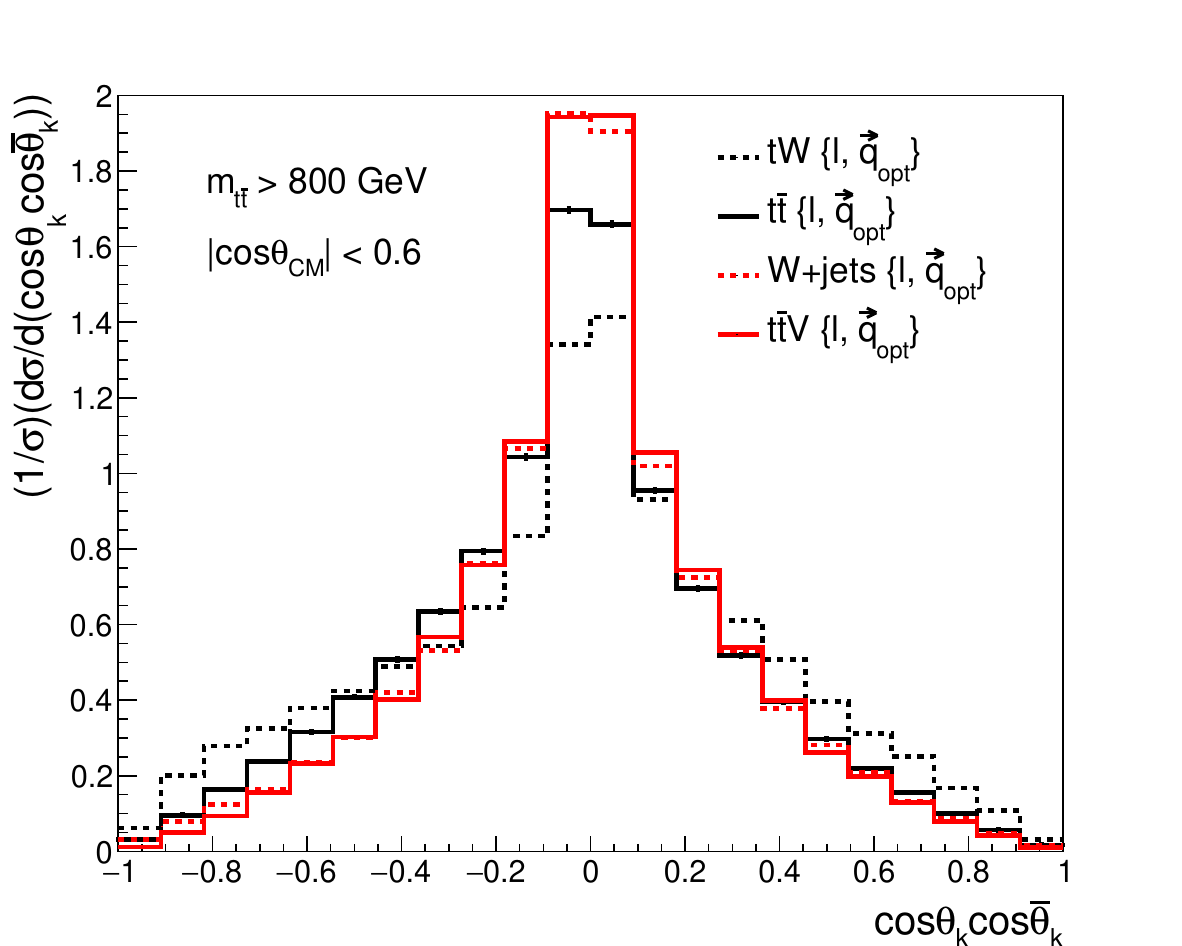}\hspace*{-0.66cm}
    \includegraphics[width=.35\textwidth]{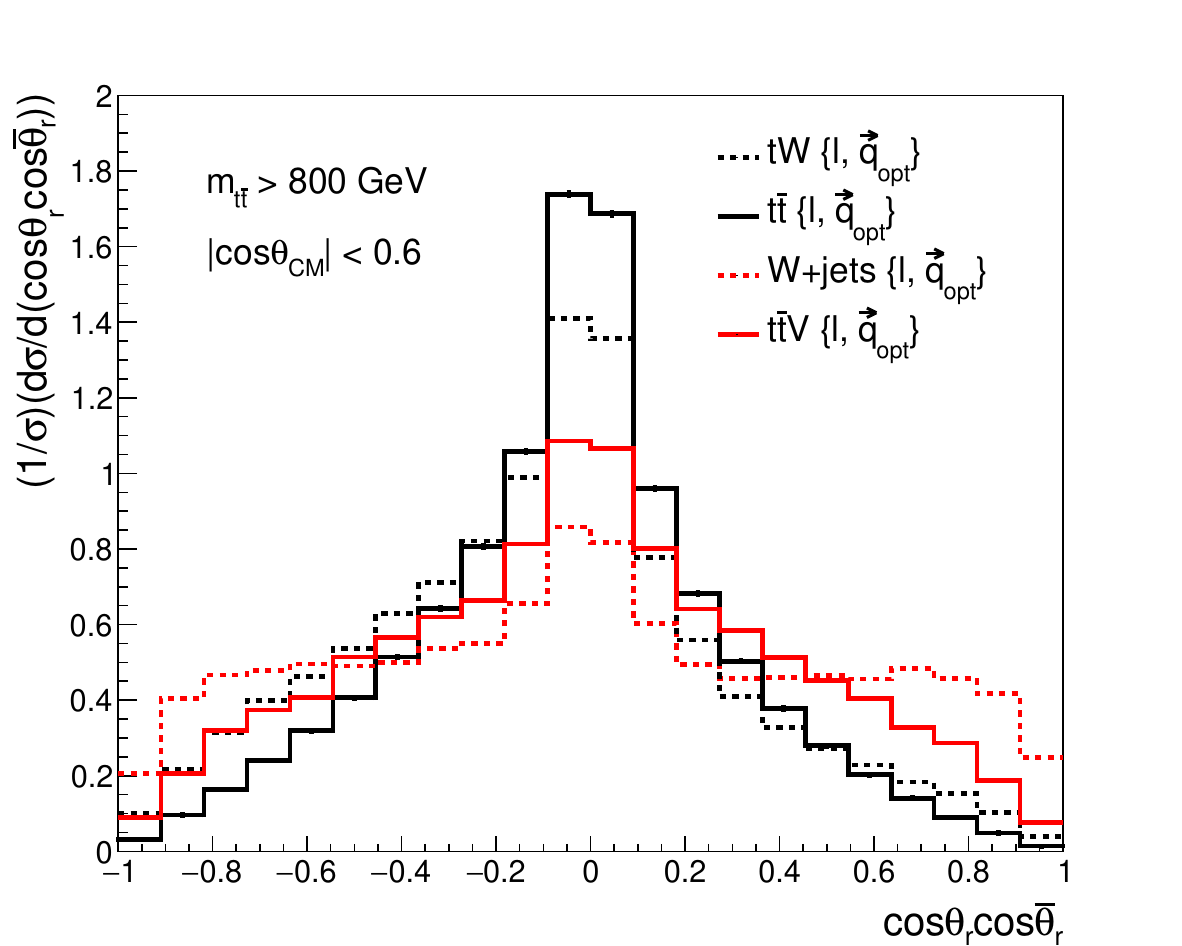}\hspace*{-0.66cm}
    \includegraphics[width=.35\textwidth]{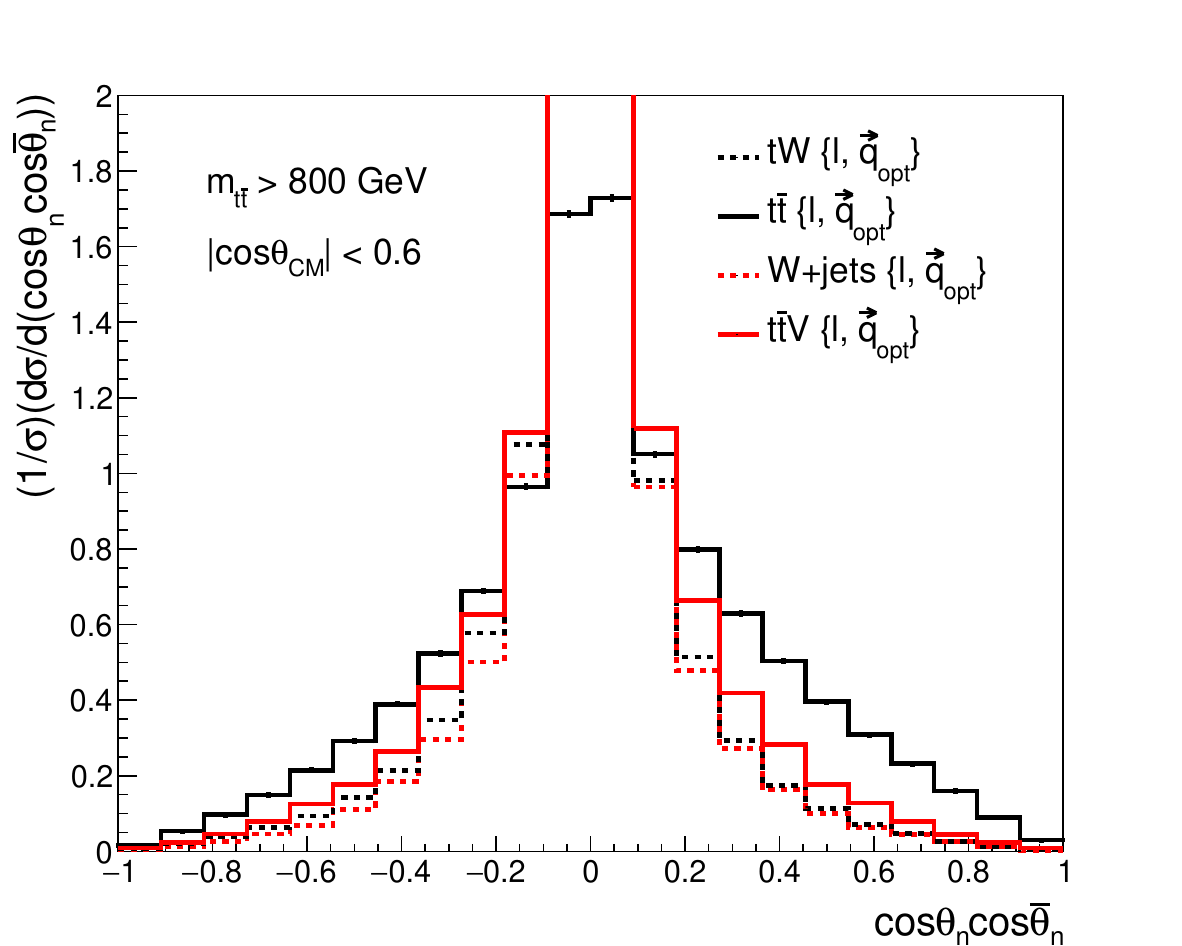}
    \caption{Differential angular distributions with respect to $\cos\theta^i_\ell \cos\bar \theta^i_{q_\text{opt}}$ for $i=k$ (left), $i=r$ (middle) and $i=n$ (right) for both signal and backgrounds. We display the parton level correlations between the charged lepton and the optimal hadronic direction. We apply the selections $m_{t\bar{t}}\geq 800$ GeV and $|\cos\theta_\text{CM}|\leq 0.6$.}
    \label{fig:cosijbkg}
\end{figure*}

Parton-level differential angular distributions for backgrounds are shown in Fig.~\ref{fig:cosijbkg}. Here, the angles are calculated between the charged lepton and the optimal hadronic direction. We also added the parton-level distribution for the signal as a reference.
%

\section{UNFOLDING ALGORITHM}
\label{app:unfold}

We start the unfolding procedure with the MC truth distribution $x^{init}$ and the corresponding MC measured distribution $b^{init}$. The response matrix $A_{ij}$  is the probability that an event generated in the true bin $j$ will be found in the measured bin $i$. The response matrix is then used to solve the system $A x = b$, where $b$ is the measured distribution (the distribution we want to unfold).
Before using SVD of the response matrix, a suitable rescaling is performed and a new distribution $w_{i}=x_{i}/x_{i}^{init}$ is introduced so that $A$ can be filled with the number of events rather than probabilities. An additional rescaling may be needed if the covariance matrix of $b$ is not diagonal, which is not the case in our analysis.

Then we perform the ridge regression (also known as Tikhonov regularization) by minimizing 
\begin{align}
(\tilde{A}w-\tilde{b})^{T}(\tilde{A}w-\tilde{b})-\tau (Cw)^{T}(Cw), 
\end{align}
where $C$ is a constant matrix that minimizes the \textit{curvature} of $w$. 
Using the SVD of $\tilde{A}C^{-1}=USV^{T}$ and introducing $d=U^{T}\tilde{b}$ and $z=V^{T}C^{-1}w$, the solution becomes $z_{i}=d_{i}/s_{i}$, where $s_{i}$ is the $i$th singular value of $\tilde{A}C^{-1}$.
The regularization parameter $\tau$ is related to the singular values of $\tilde{A}C^{-1}$. We find the best choice of $\tau$ from the $\log d_{i}$ distribution, which is the coefficient of the decomposition of $\tilde{b}$ in a basis defined by the $i$th column of $U$. Usually, only the first $m$ terms are statistically significant. The best choice of this parameter is then $\tau = s_{m}^{2}$, \emph{i.e.}, the square of the $m$th singular value~\cite{H_cker_1996}.  

To compute the $C_{ij}$ coefficients defined in Eq.~\eqref{eq:cij} we define the asymmetries as 
\begin{align}
    A = \frac{\sum_{\text{bins}>0}\,N_{i}-\sum_{\text{bins}<0}\,N_{i}}{\sum_{\text{all bins}}\,N_{i}}\,,
\end{align}
where $N_{i}$ is the content of bin $i$ and $\text{bins}>0$ ($\text{bins}<0$) denotes the number of bins with $\cos\theta^{i}_\ell \cos\bar{\theta}^{i}_{q_\text{opt}}>0$ ($\cos\theta^{i}_\ell \cos\bar{\theta}^{i}_{q_\text{opt}}<0$). Using error propagation we compute the error in the asymmetry as 
\begin{align}
    \sigma_{A} =\sqrt{ \sum_{i,j} g_{i}g_{j}V_{ij}}\,,
\label{eq:errorAsymm}
\end{align}
where $V$ is the covariance matrix of the unfolded $\cos\theta^{i}_\ell \cos\bar{\theta}^{i}_{q_\text{opt}}$ distribution and $g_{i}$ is a factor defined below 
\begin{align}
        g_{i} = \begin{dcases}
		\frac{-2\sum_{\text{bins}>0}\,N_{i}}{(\sum_{\text{all bins}}\,N_{i})^2}\,, & i\in \qq*{bins}\!\!\!\!\!<0 \\	
	    \frac{2\sum_{\text{bins}<0}\,N_{i}}{(\sum_{\text{all bins}}\,N_{i})^2}\,, & i\in \qq*{bins}\!\!\!\!\!>0 \\			
    \end{dcases}\,.
\end{align}

Note that Eq.~\eqref{eq:errorAsymm} accounts for all the bin-to-bin correlations that the unfolding might have introduced, as it uses the full covariance matrix of the unfolded distribution.    

In Fig.~\ref{fig:cosicosj}, we illustrate the robustness of the unfolding algorithm, applying it to the distributions $\cos\theta^{k}_\ell \cos\bar{\theta}^{k}_{q_\text{opt}}$ and $\cos\theta^{r}_\ell \cos\bar{\theta}^{r}_{q_\text{opt}}$  with 22 bins and $m=6$, as well as  $\cos\theta^{n}_\ell \cos\bar{\theta}^{n}_{q_\text{opt}}$ with 22 bins and $m=7$ in the signal region for entanglement. However, for the final results, we choose 2 bins and $m=2$ to unfold all distributions, as we are interested only in the regions with $\cos\theta^{i}_\ell \cos\bar{\theta}^{i}_{q_\text{opt}}>0$ and $\cos\theta^{i}_\ell \cos\bar{\theta}^{i}_{q_\text{opt}}<0$. Our analysis accounts for both the statistical error stemming from the measured distribution and the MC error in the response matrix, calculated with 100 pseudo-experiments.


\section{SPACELIKE SEPARATION}
\label{app:spacelike}

To ensure that the top quarks are spacelike separated when the spin information is passed on to the  decay products, they have to be relatively apart from one another so that no information can be passed in between when they decay~\cite{Severi:2021cnj}. In the center of the mass frame of $t\bar{t}$, the distance between their decay locations is given by $(t_1+t_2)v$, where $t_1$ and $t_2$ are the decay times of the top and anti-top quarks, and $v$ is the magnitude of their velocity. The maximum distance that information can travel between their decay times is given by $|t_1-t_2|c$, where $c$ is the speed of the light. Thus, spacelike separation requires the following inequality:
\begin{equation}\label{eq:INeq}
    \dfrac{|t_1-t_2|}{t_1+t_2}< \frac{v}{c}  = \sqrt{1 - \frac{4 m_t^2}{m_{t\bar t}^2}}.
\end{equation}
Counting the fraction of events where this inequality holds tells us how often the tops are spacelike separated.

\section{LOOPHOLES}
\label{app:loopholes}

Measurement of a loophole-free Bell violation is a more intricate task compared to probing entanglement. The LHC, not originally designed for testing Bell inequalities, can only explore weak violations. It is noteworthy that preparing  loophole-free setups took decades, with the first measurement occurring around 2015~\cite{Hensen:2015ccp,PhysRevLett.115.250401,PhysRevLett.115.250402}, recognized by the 2022 Physics Nobel Prize.  At the LHC, a significant challenge is the inability to apply external intervention for freely choosing the orientation of the measurement axis for the top pair, which poses a challenge to the free-will loophole. Additionally, achieving causal independence of the decays is only possible at a statistical level~\cite{Severi:2021cnj}. Last, the LHC analyses focus on a subset of events, introducing the detection loophole.

\bibliography{references}

\end{document}